# Beyond Procedural Compliance: Human Oversight as a Dimension of Well-being Efficacy in AI Governance

**Yao Xie**
School of Medicine, University College Dublin
Dublin, Ireland
`toyaoxie@gmail.com`

**Walter Cullen**
School of Medicine, University College Dublin
Dublin, Ireland
`walter.cullen@ucd.ie`

## Abstract

Major AI ethics guidelines and laws, including the EU AI Act, call for effective human oversight, but do not define it as a distinct and developable capacity. This paper introduces human oversight as a well-being capacity, situated within the emerging Well-being Efficacy framework. The concept integrates AI literacy, ethical discernment, and awareness of human needs, acknowledging that some needs may be conflicting or harmful. Because people inevitably project desires, fears, and interests into AI systems, oversight requires the competence to examine and, when necessary, restrain problematic demands.

The authors argue that the sustainable and cost-effective development of this capacity depends on its integration into education at every level, from professional training to lifelong learning. The frame of human oversight as a well-being capacity provides a practical path from high-level regulatory goals to the continuous cultivation of human agency and responsibility essential for safe and ethical AI. The paper establishes a theoretical foundation for future research on the pedagogical implementation and empirical validation of well-being effectiveness in multiple contexts.

## 1 Introduction

The contemporary world is entering an era defined by artificial intelligence (AI) (Xu et al., 2024). Driven by rapid innovation, AI transforms how people learn, work, communicate, and decide (Abuzaid, 2024; Afroogh et al., 2024). This movement marks an irreversible shift toward AI-mediated environments where intelligent systems increasingly shape everyday life and well-being(Singh & Tholia, 2024). The transformation brings new opportunities for creativity and efficiency but also exposes people to losses in agency, coherence, and collective trust. The speed of technological change, the weakening of human voice, and public indifference toward data protection reveal how easily individuals trade agency for convenience (Yatani et al., 2024). These conditions make human oversight not only a technical necessity but also a fundamental capacity for collective well-being (Corrêa et al., 2025; Langer et al., 2024; Sterz et al.).

Human oversight has become a central principle of global AI governance (Koulu, 2020). The European Union AI Act, the OECD AI Principles, and the UNESCO Recommendation on the Ethics of Artificial Intelligence all emphasise its role in keeping technology aligned with human priorities (Enqvist, 2023). These frameworks recognise that AI should not operate without human judgment.

However, their practical interpretations often focus on institutional examples: a clinician approving an algorithmic diagnosis, a financial analyst authorising an automated transaction, or a teacher monitoring AI-assisted assessment (Koulu, 2020). These examples describe regulated professional contexts but fail to reflect the broader scope of AI influence such as the subtle ways in which design and daily use shape human experience (Koulu, 2020).

Most people encounter AI in ordinary contexts rather than in professional ones. They interact with recommendation systems that shape preferences, social media feeds that influence belief, and digital platforms that collect and reuse personal data in what can be described as *hybrid space*— 'refer to merging physical and digital spaces'(de Souza e Silva et al., 2025). Public institutions also increasingly depend on algorithmic tools in police, welfare allocation, and recruitment. In such environments, oversight becomes diffuse and largely invisible (Koulu, 2020; Sterz et al.). Current regulations seem to assume that people already have the awareness and critical discernment needed to make informed choices. In reality, many people engage with AI outputs passively, lacking the reflective capacity necessary for effective oversight (Chen et al., 2023). Humans also have inherent limitations in decision making, often relying on cognitive shortcuts that reduce awareness and deliberation (Curran, 2015; Dale, 2015; Yoder & Decety, 2018).

The governance challenge extends beyond technical regulation to the cultivation of human awareness and agency across all levels of society (Sigfrids et al., 2023; Van Popering-Verkerk et al., 2022). Governance cannot depend solely on external rules or audits, it requires stronger human capacity (Enqvist, 2023; Yeung et al., 2020). As AI becomes embedded in daily life, the question is not how a single person supervises a specific system, but how human agency can be maintained in complex, distributed, and rapidly evolving environments. This challenge calls for scalable human oversight—the capacity to preserve human judgment and ethical control when direct supervision of each decision cannot be achieved (Green, 2022).

The authors argue that effective scalable oversight depends on the cultivation of what we term well-being efficacy—the integrated human capacity to sustain agency, coherence, and ethical clarity within complex and changing contexts (Welzel & Inglehart, 2010). Human oversight represents a critical application of this broader capacity in AI governance. This paper presents a theoretical framework that repositions oversight from a compliance mechanism to a developable human capacity. Although the comprehensive well-being efficacy framework, including its dimensional structure and measurement instruments, is under ongoing empirical development (Anonymity et al., in preparation), this paper focuses specifically on establishing the conceptual foundation for its application to AI governance. This reframing addresses a critical gap: current regulations assume oversight capability exists while providing no pathway for its systematic development. Sustainable governance emerges not through structural redesign alone but through the continuous cultivation of human capacity across education, professional training, and lifelong learning.

## 2   Redefining Human Oversight

Human oversight refers to the capacity to monitor, interpret, and intervene when required in the operation of complex systems. Traditional frameworks describe this through models such as human in the loop, human on the loop, or human in command, which define specific points of human involvement in technical processes (Enarsson et al., 2022; Tschiatschek et al., 2024; Wu et al., 2022). These configurations remain useful in regulated institutional contexts but represent only partial expressions of oversight. Oversight extends beyond procedural checkpoints (Enqvist, 2023; Sterz et al.; Wu et al., 2022). It depends on human motives, awareness, and the ability to recognise when reflection or ethical judgment is required (Greene et al., 2008). The quality of oversight depends not only on a system's formal risk level but also on human capacity to identify moments that call for reflection, restraint, or ethical consideration (Tschiatschek et al., 2024). Oversight is more than a compliance mechanism. It represents an adaptive function of consciousness and moral reasoning that operates across diverse contexts and scales (Conitzer et al., 2024; Guzman et al., 2022).

Oversight also functions in both directions. Human choices, data contributions, and patterns of interaction influence how AI systems learn and adapt (Melmed, 2020). Because people inevitably project desires, fears, and interests into technological systems, oversight requires the competence to examine and, when necessary, restrain problematic demands. Awareness of this reciprocal influence transforms oversight from reactive supervision into active participation in shaping technological development. Every digital decision, whether through clicking, sharing, consenting, or remaining silent, creates the informational environment that shapes collective experience (Callaghan, 2018; Conitzer et al., 2024; Xie et al., 2025). Oversight in this broader sense recognises human agency not only in evaluating AI outputs but also in shaping the conditions under which these systems evolve (Conitzer et al., 2024)

## 2.1 From Individual Capacity to Distributed Social Function

In settings where direct supervision of every algorithmic decision is nearly impossible (Sudhir et al.), human oversight must function as a distributed social capacity rather than a form of concentrated institutional control (Kenton et al., 2024). This need for scalability reflects not only a technical challenge of coverage and efficiency but also the demand for humans to extend awareness, discernment, and responsibility across AI mediated contexts (Sterz et al.). Scalable oversight arises when reflective and ethical capacities are diffused throughout society, allowing individuals in different roles to recognise critical moments that require attention or intervention. This distribution does not depend on identical expertise. It creates a shared foundation of discernment that enables people to identify when systems exceed reasonable boundaries, when outcomes need scrutiny, or when human values are placed at risk.

This approach supports sustainable governance because developed human capacities can adapt to changing technologies without constant structural redesign (Orr & Burch, 2025). Scalable oversight invests in human judgment that retains relevance across technological generations, offering advantages over systems that require continuous updating to match technical evolution (Anadon et al., 2016). The governance challenge thus shifts from designing increasingly complex external controls to cultivating the human capacities that enable responsible engagement with technology across diverse and changing contexts (Cosens et al., 2021).

## 3 Well-being Efficacy as the Foundation for Scalable Oversight

Well-being efficacy provides the conceptual foundation for this distributed oversight function. It refers to the integrated human capacity to preserve agency, coherence, and ethical clarity within complex and changing contexts. Well-being efficacy combines cognitive, emotional, and relational dimensions. Metacognitive awareness supports reflection on one's own thinking and judgment. Epistemic vigilance enables critical evaluation of information sources and claims (Bielik & Krell, 2025). Systems thinking encourages recognition of interconnections and long term consequences. Shared responsibility strengthens collective orientation toward common well-being(Cosens et al., 2021; Van Popering-Verkerk et al., 2022). Emerging evidence from ongoing research indicates that these capacities can be further organised into a structured set of interrelated dimensions that explain how humans maintain agency and ethical clarity in AI mediated environments.

When individuals develop these capacities, they contribute to a collective oversight function operating across multiple scales and contexts (Orr & Burch, 2025). This is not narrow technical expertise but a form of practical wisdom that enables people to pause, question, contextualise, and act with moral clarity in technologically mediated environments.

This distributed capacity appears through complementary mechanisms spanning technical, organisational, social, and educational domains (Parthasarathy et al., 2024; Sigfrids et al., 2023). Technical transparency tools such as model cards and interpretability interfaces provide information that becomes meaningful oversight only when used with reflective awareness (Kalodanis et al.). Organisational practices including participatory design and red teaming formalise stakeholder involvement, though their effectiveness depends on participants' capacity for critical reflection. Social structures such as citizen observatories and public deliberation create spaces for collective evaluation, but these mechanisms function effectively only when participants have the capacity for thoughtful engagement (Parthasarathy et al., 2024; Ter-Minassian, 2025). Educational systems complete this architecture through the cultivation of evaluative judgment needed for meaningful participation across these domains (Ter-Minassian, 2025).

The relationship between individual capacity and collective oversight strengthens through mutual reinforcement. Education develops the capacities required for genuine engagement in oversight, and active participation in oversight processes deepens understanding and discernment. This reciprocal development enables oversight to evolve with technology, maintaining governance through human growth instead of static regulatory design (Sigfrids et al., 2023).

### 3.1 Human Oversight as the Practice of Well-being Efficacy

Human oversight, properly understood, expresses well-being efficacy in relation to technological systems (Schiff et al.). It extends beyond technical comprehension to include the ability to pause before acting, to examine underlying motives and values, and to intervene with intention when systems deviate from human priorities (Hille et al., 2023). Current approaches to AI literacy mainly improve understanding of how AI systems function and how they can be used responsibly. This informational literacy is valuable but limited. It equips individuals to operate within existing systems but does not enable them to question, evaluate, or ethically redirect AI behaviour as systems evolve or influence human judgment. Cultivating human oversight as an expression of well-being efficacy extends from comprehension to conscious regulation. It develops meta awareness, ethical discernment, and reflective self-control, allowing individuals to preserve autonomy and moral clarity within dynamic technological environments. AI literacy enhances knowledge about technology, whereas well-being efficacy strengthens the internal capacity to think, reflect, and make value based decisions within it.

Oversight in this sense supports inner coherence and helps people navigate technological complexity without losing autonomy or moral clarity. It anchors human judgment within environments that accelerate perception and decision making, creating space for ethical consideration even when systems operate faster than biological cognition. Developing this capacity also strengthens healthy boundaries between humans and technology. As AI becomes integrated across economic, social, and ecological systems, it influences individual choices, collective norms, institutional behaviour, and environmental decision making. The capacity to determine when to depend on or resist technological intervention is central to maintaining human agency. Oversight as an expression of well-being efficacy enhances the ability to pause, reflect, and make deliberate judgments across diverse contexts instead of reacting automatically to technological convenience or pressure.

Every decision, whether made directly by humans or mediated through algorithms, carries consequences for human well-being (Schiff et al.). Well-being in this context encompasses collective conditions for coexistence, trust, and moral integrity. AI was created to enhance the human condition, though its contribution to that goal depends entirely on human discernment and intentionality. Technology itself has no intrinsic moral direction. It amplifies human purposes, whether constructive or destructive.

### 3.2 Evolutionary and Cognitive Adaptation

The cultivation of well-being efficacy responds to the growing mismatch between biological evolution and technological acceleration. Human cognition developed in environments that offered gradual feedback and tangible consequences(Gonzalez-Cabrera, 2020; van Vugt et al., 2024). Information once moved at a pace that allowed reflection, verification, and moral consideration. AI now mediates perception, communication, and decision-making at scales and speeds far beyond the conditions for which the human mind evolved (Chen et al., 2023). Information arrives faster than most people can process, and decisions occur within systems that exceed direct human comprehension (Schmitt et al., 2021). This creates a structural gap between the rate of technological change and the human capacity to regulate its effects (van Vugt et al., 2024).

The imbalance weakens epistemic vigilance, the ability to assess the reliability of information, and produces cognitive overload, emotional fatigue, and social apathy(Schmitt et al., 2021). Many people experience a loss of agency and voice in digital environments (Couldry & Mejias, 2019). Indifference toward personal data protection exemplifies this problem. Individuals may recognise risk but often feel powerless to act. Developing oversight as a dimension of well-being efficacy functions as a conscious adaptation that restores balance between human awareness and the accelerating technological context. Strengthening oversight re-establishes reflection, discernment, and participation in shaping the systems that influence collective life.

### 3.3 Oversight and Well-being Efficacy in Dynamic Contexts

Well-being efficacy extends beyond deciding whether to accept or reject technology. It embodies the capacity to maintain coherence, direction, and meaning within changing environments. It enables individuals and organisations to orient themselves amid uncertainty, identify what matters, and act with integrity (Welzel & Inglehart, 2010). Oversight in this sense integrates adaptation with reflection. It helps people remain centred instead of being absorbed by the speed and opacity of technological systems.

This capacity reinforces ethical and relational awareness. It supports empathy, dialogue, and shared purpose even when communication is technologically mediated (Floridi, 2023). Human oversight as a practice of well-being efficacy strengthens decision-making about where and how AI should be applied. As AI becomes embedded in health, education, governance, and environmental management, decisions involve not only technical feasibility but also moral suitability. The ability to discern when AI use is beneficial, when it weakens human judgement or social trust, and when mixed human–machine approaches are most appropriate, becomes essential. In this way, oversight evolves from a reactive safeguard into a proactive guide for responsible innovation. It helps societies maintain continuity and collective direction in periods of rapid transformation, ensuring that AI adoption contributes to human well-being and keeps human judgement at the centre of decision-making.

## 4 Educational Implementation for Sustainable Development

The accelerating development of AI makes the cultivation of well-being efficacy through education an urgent necessity (UNESCO, 2021; Long & Magerko, 2020). As technologies advance faster than institutional responses, reliance on regulation or technical auditing becomes costly and unsustainable (Floridi, 2021). Safeguards require constant revision, but without parallel growth in human awareness, ethical reasoning, and reflective judgement, their effectiveness soon declines. Building human capacity offers a more durable and economical foundation for responsible oversight. It enables adaptation to new technologies without continual structural reform and strengthens collective resilience (Seldon & Abidoye, 2018). Education provides the most stable means of renewing these capacities across generations and professional fields.

Implementation can occur at several levels. In general education, teaching about how digital systems shape perception and behaviour fosters reflective and ethical participation (Ng, 2021). Students can learn to question credibility, context, and motive to support critical thinking (Paul & Elder, 2014). In professional education, fields such as healthcare, teaching, and public administration can combine AI literacy with ethical reasoning through scenario-based learning and deliberative exercises (Popenici & Kerr, 2017). Lifelong learning programmes can maintain this growth for adults who must update oversight practices as technologies and norms evolve (Merriam & Baumgartner, 2020).

Although few initiatives explicitly teach oversight, existing domains such as critical thinking, media literacy, and ethics offer strong starting points (Buckingham, 2019). Integrating these subjects into curricula embeds awareness and moral reasoning in everyday practice. Investment in education reduces dependence on external auditing and reinforces collective responsibility. It strengthens agency, reflection, and cooperation within organisations and communities (Illeris, 2018). When people learn to pause, evaluate, and act with discernment, societies become more adaptive and less vulnerable to manipulation (Serrat, 2017). Education transforms oversight from a procedural requirement into a cultural habit. It develops the psychological, ethical, and social foundations that allow humans to remain self-aware and value-driven as technology evolves. Embedding well-being efficacy in both formal and informal learning ensures that technological progress advances human priorities and remains a shared, sustainable enterprise (United Nations, 2015).

## 5 Conclusion and Discussion

Human oversight is often treated as a procedural safeguard within AI governance, but its deeper significance lies in the growth of human capacity itself. Understanding oversight as an expression of well-being efficacy transforms its aim from control to cultivation and from regulation to the continuous exercise of awareness and ethical reasoning. Effective oversight depends on cognitive and moral maturity that enables reflection and value-based action within complex systems.

AI was designed to enhance human life, but this purpose can only be achieved through human agency. Technology cannot ensure positive outcomes without guidance from people who possess discernment and responsibility. Oversight therefore represents both a skill and a moral commitment to align technological development with human purpose. The capacities that constitute well-being efficacy extend beyond AI. They strengthen empathy, cooperation, and thoughtful judgement across all areas of society. Every decision, whether mediated through technology or made directly, influences collective well-being. This continuity affirms a simple principle: decisions gain meaning through their consequences for human life.

Viewing oversight through the lens of well-being efficacy offers a coherent and sustainable model of governance. It links institutional structures with personal growth and anchors technological progress in human understanding. Education and lifelong learning secure the moral and psychological foundations needed for innovation to serve collective well-being. Through this lens, oversight becomes an enduring human capacity that supports ethical agency, informed choice, and resilience within an evolving technological world. It reflects the continuing expression of consciousness, ethics, and purpose in an accelerating age.

## 5.1 Theoretical and Practical Implications

The well-being efficacy framework positions human oversight within a broader understanding of human agency in technological contexts. Although this paper concentrates on AI governance, the underlying construct connects to a wider inquiry into how humans maintain autonomous functioning in conditions of accelerating technological change. The framework's dimensional structure, measurement approaches, and empirical validation are the focus of ongoing research (Xie et al., in preparation).

From a policy perspective, the framework suggests that effective AI regulation must extend beyond technical standards and procedural compliance to encompass sustained investment in human capacity development. International governance initiatives, including the *UNESCO Recommendation on the Ethics of Artificial Intelligence* (UNESCO, 2021) and the *EU AI Act* (Corrêa et al., 2025; Enqvist, 2023), emphasise human oversight but rarely define how such capacity can be cultivated. The framework therefore introduces a means of linking regulatory aims with human development goals, aligning with broader sustainable governance approaches that integrate education and capacity building as part of societal resilience (Anadon et al., 2016; Orr & Burch, 2025; Van Popering-Verkerk et al., 2022).

From an educational perspective, the framework implies that AI literacy should expand from technical comprehension toward capacity cultivation. Knowledge of how AI systems operate is necessary but insufficient for sustaining human judgement. Educational programmes should focus on metacognitive awareness, ethical discernment, and relational sensitivity that enable humans to maintain agency and coherence in AI-mediated environments (Buckingham, 2019; Illeris, 2018; Long & Magerko, 2020; Merriam & Baumgartner, 2020; Ng, 2021; Paul & Elder, 2014). Embedding these dimensions within general, professional, and lifelong learning can help ensure that education functions as a durable mechanism of ethical and cognitive adaptation (Seldon & Abidoye, 2018; Popenici & Kerr, 2017).

From an organisational perspective, effective AI deployment requires that human capacity and technical infrastructure evolve in parallel. Institutions should integrate reflective training and participatory mechanisms that enhance collective discernment and ethical awareness (Parthasarathy et al., 2024; Sigfrids et al., 2023). This investment supports long-term adaptability and reduces dependence on costly external oversight, reinforcing a culture of shared responsibility within complex sociotechnical systems (Cosens et al., 2021; Serrat, 2017).

## 5.2 Limitations

This paper offers a theoretical foundation for understanding the relationship between human oversight and well-being efficacy. Its purpose is conceptual clarification rather than the presentation of empirical results. The framework outlines how human capacities underpin scalable oversight, yet the mechanisms through which these capacities develop and interact across contexts require empirical study. Further research should examine these relationships through mixed-method approaches that combine qualitative interpretation with psychometric validation (Bielik & Krell, 2025).

Measurement remains a critical challenge. Operational tools for assessing well-being efficacy and tracking its growth through educational or organisational interventions are still in development. Establishing valid indicators will be essential to translate the framework into applied practice and to evaluate its outcomes across cultural settings. Cultural variability represents another limitation. The framework draws primarily on Western philosophical and policy traditions that emphasise autonomy and individual agency (Welzel & Inglehart, 2010; Yeung et al., 2020). Its relevance across non-Western contexts, where understandings of agency, responsibility, and collective oversight may differ, warrants further exploration.

The implementation of educational and policy transformation also involves considerable complexity. Structural reform requires alignment across political, institutional, and economic systems, as well as collaboration between educators, policymakers, and communities (Orr & Burch, 2025). These challenges extend beyond the theoretical focus of the present paper but remain important for translating the framework into practice.

### 5.3 Future Directions

This study lays the conceptual foundation for understanding human oversight as a dimension of well-being efficacy. The current focus is to refine the theoretical structure, scope, and philosophical coherence of the construct. Ongoing work examines educational design and pedagogical models that cultivate well-being efficacy in practice, with attention to curriculum development and capacity building. The next stage will develop measurement approaches and policy evaluation frameworks that translate the construct into operational tools for institutions and communities. These lines of research aim to connect conceptual understanding with implementation and create a coherent path from theory to practice that supports sustainable AI governance through the cultivation of human capacity.

## Acknowledgments

Removed for the purpose of peer review.

## References

Abuzaid, A. N. (2024, April 18–19). *Strategic AI integration: Examining the role of artificial intelligence in corporate decision-making.* 2024 International Conference on Knowledge Engineering and Communication Systems (ICKECS).

Afroogh, S., Akbari, A., Malone, E., Kargar, M., & Alambeigi, H. (2024). Trust in AI: Progress, challenges, and future directions. *Humanities and Social Sciences Communications, 11*(1), 1–30.

Anadon, L. D., Chan, G., Harley, A. G., Matus, K., Moon, S., Murthy, S. L., & Clark, W. C. (2016). Making technological innovation work for sustainable development. *Proceedings of the National Academy of Sciences, 113*(35), 9682–9690. https://doi.org/10.1073/pnas.1525004113

Bielik, T., & Krell, M. (2025). Developing and evaluating the extended epistemic vigilance framework. *Journal of Research in Science Teaching, 62*(3), 869–895.

Buckingham, D. (2019). *Teaching media in a digital age: Learning, literacy, and culture.* Polity Press.

Callaghan, C. W. (2018). Surviving a technological future: Technological proliferation and modes of discovery. *Futures, 104*, 100–116. https://doi.org/10.1016/j.futures.2018.08.001

Chen, V., Liao, Q. V., Wortman Vaughan, J., & Bansal, G. (2023). Understanding the role of human intuition on reliance in human-AI decision-making with explanations. *Proceedings of the ACM on Human-Computer Interaction, 7*(CSCW2), 1–32.

Conitzer, V., Freedman, R., Heitzig, J., Holliday, W. H., Jacobs, B. M., Lambert, N., Mossé, M., Pacuit, E., Russell, S., & Schoelkopf, H. (2024). Social choice should guide AI alignment in dealing with diverse human feedback. *arXiv preprint arXiv:2404.10271*.

Corrêa, A. M., Garsia, S., & Elbi, A. (2025). Better together? Human oversight as means to achieve fairness in the European AI Act governance. *Cambridge Forum on AI: Law and Governance, 1*, e29. https://doi.org/10.1017/cfl.2025.10010

Cosens, B., Ruhl, J. B., Soininen, N., Gunderson, L., Belinskij, A., Blenckner, T., Camacho, A. E., Chaffin, B. C., Craig, R. K., Doremus, H., Glicksman, R., Heiskanen, A. S., Larson, R., & Simila, J. (2021). Governing complexity: Integrating science, governance, and law to manage accelerating change in the globalized commons. *Proceedings of the National Academy of Sciences, 118*(36). https://doi.org/10.1073/pnas.2102798118

Couldry, N., & Mejias, U. A. (2019). *The costs of connection: How data is colonizing human life and appropriating it for capitalism.* Stanford University Press.

Curran, E. T. (2015). Outbreak column 16: Cognitive errors in outbreak decision making. *Journal of Infection Prevention, 16*(1), 32–38. https://doi.org/10.1177/1757177414562057

Dale, S. (2015). Heuristics and biases: The science of decision-making. *Business Information Review, 32*(2), 93–99.

de Souza e Silva, A., Campbell, S. W., & Ling, R. (2025). Hybrid space revisited: From concept toward theory. *Communication Theory, 35*(1), 14–24. https://doi.org/10.1093/ct/qtae023

Enarsson, T., Enqvist, L., & Naarttijärvi, M. (2022). Approaching the human in the loop: Legal perspectives on hybrid human/algorithmic decision-making in three contexts. *Information & Communications Technology Law, 31*(1), 123–153.

Enqvist, L. (2023). 'Human oversight' in the EU Artificial Intelligence Act: What, when, and by whom? *Law, Innovation and Technology, 15*(2), 508–535. https://doi.org/10.1080/17579961.2023.2245683

Fachner, J., & Witte, T. (2018). Human decision-making in high-stakes contexts: Psychological mechanisms and legal implications. *Psychology, Crime & Law, 24*(3), 279–295. https://doi.org/10.1080/1068316X.2017.1414817

Floridi, L. (2021). The ethics of artificial intelligence: Principles, challenges, and opportunities. *Philosophy & Technology, 34*(1), 1–13. https://doi.org/10.1007/s13347-021-00460-6

Floridi, L. (2023). *The ethics of artificial intelligence: Principles, challenges, and opportunities.* Oxford University Press. https://doi.org/10.1093/oso/9780198883098.001.0001

Gonzalez-Cabrera, I. (2020). [Review of *Becoming Human: A Theory of Ontogeny*, by M. Tomasello]. *History and Philosophy of the Life Sciences, 42*(4), 1–5. http://www.jstor.org/stable/45410939

Green, B. (2022). The flaws of policies requiring human oversight of government algorithms. *Computer Law & Security Review, 45*, 105681. https://doi.org/10.1016/j.clsr.2022.105681

Greene, J. D., Morelli, S. A., Lowenberg, K., Nystrom, L. E., & Cohen, J. D. (2008). Cognitive load selectively interferes with utilitarian moral judgment. *Cognition, 107*(3), 1144–1154. https://doi.org/10.1016/j.cognition.2007.11.004

Guzman, R. A., Barbato, M. T., Sznycer, D., & Cosmides, L. (2022). A moral trade-off system produces intuitive judgments that are rational and coherent and strike a balance between conflicting moral values. *Proceedings of the National Academy of Sciences, 119*(42), e2214005119. https://doi.org/10.1073/pnas.2214005119

Hille, E. M., Hummel, P., & Braun, M. (2023). Meaningful human control over AI for health? A review. *Journal of Medical Ethics.*

Illeris, K. (2018). *Contemporary theories of learning: Learning theorists. . . in their own words* (2nd ed.). Routledge.

Kalodanis, K., Rizomiliotis, P., Feretzakis, G., Kalles, D., Verykios, V., & Anagnostopoulos, D. (2024). Enhancing transparency in large language models to meet EU AI Act requirements. *arXiv preprint arXiv:2405.10706.*

Kenton, Z., Siegel, N., Kramár, J., Brown-Cohen, J., Albanie, S., Bulian, J., Agarwal, R., Lindner, D., Tang, Y., & Goodman, N. (2024). On scalable oversight with weak LLMs judging strong LLMs. *Advances in Neural Information Processing Systems, 37*, 75229–75276.

Koulu, R. (2020). Proceduralizing control and discretion: Human oversight in artificial intelligence policy. *Maastricht Journal of European and Comparative Law, 27*(6), 720–735.

Langer, M., Baum, K., & Schlicker, N. (2024). Effective human oversight of AI-based systems: A signal detection perspective on the detection of inaccurate and unfair outputs. *Minds and Machines, 35*(1), 1. https://doi.org/10.1007/s11023-024-09701-0

Long, D., & Magerko, B. (2020). What is AI literacy? Competencies and design considerations. *Proceedings of the 2020 CHI Conference on Human Factors in Computing Systems*, 1–16. https://doi.org/10.1145/3313831.3376727

Melmed, M. L. (2020). Bound by infinities: Technology, immediacy, and our environmental crisis. *American Journal of Psychoanalysis, 80*(3), 342–353. https://doi.org/10.1057/s11231-020-09258-8

Merriam, S. B., & Baumgartner, L. M. (2020). *Learning in adulthood: A comprehensive guide* (4th ed.). Wiley.

Ng, W. (2021). *Critical digital literacy: Dispositions and skills for the 21st century learner.* Springer.

Orr, C. J., & Burch, S. (2025). Transformative capacities for navigating system change: A framework for sustainability research and practice. *Sustainability Science, 20*(3), 975–992. https://doi.org/10.1007/s11625-025-01658-y

Parthasarathy, A., Phalnikar, A., Jauhar, A., Somayajula, D., Krishnan, G. S., & Ravindran, B. (2024). Participatory approaches in AI development and governance: A principled approach. *arXiv preprint arXiv:2407.13100*.

Paul, R., & Elder, L. (2014). *The miniature guide to critical thinking concepts and tools* (7th ed.). Foundation for Critical Thinking.

Popenici, S. A. D., & Kerr, S. (2017). Exploring the impact of artificial intelligence on teaching and learning in higher education. *Research and Practice in Technology Enhanced Learning, 12*(22). https://doi.org/10.1186/s41039-017-0062-8

Schiff, D., Ayesh, A., Musikanski, L., & Havens, J. C. (2020). IEEE 7010: A new standard for assessing the well-being implications of artificial intelligence.

Schmitt, J. B., Breuer, J., & Wulf, T. (2021). From cognitive overload to digital detox: Psychological implications of telework during the COVID-19 pandemic. *Computers in Human Behavior, 124*, 106899. https://doi.org/10.1016/j.chb.2021.106899

Seldon, A., & Abidoye, O. (2018). *The fourth education revolution: Will artificial intelligence liberate or infantilise humanity?* University of Buckingham Press.

Serrat, O. (2017). *Knowledge solutions: Tools, methods, and approaches to drive organisational performance*. Springer.

Sigfrids, A., Leikas, J., Salo-Pöntinen, H., & Koskimies, E. (2023). Human-centricity in AI governance: A systemic approach. *Frontiers in Artificial Intelligence, 6*, 976887.

Singh, A., & Tholia, S. (2024). Toward the symbiocene through artificial intelligence. *AI & Society, 39*(2), 805–806. https://doi.org/10.1007/s00146-022-01546-4

Sterz, S., Baum, K., Biewer, S., Hermanns, H., Lauber-Rönsberg, A., Meinel, P., & Langer, M. (2024). On the quest for effectiveness in human oversight: Interdisciplinary perspectives.

Sudhir, A. P., Kaunismaa, J., & Panickssery, A. (2025). A benchmark for scalable oversight mechanisms.

Ter-Minassian, L. (2025). Democratizing AI governance: Balancing expertise and public participation. *arXiv preprint arXiv:2502.08651*.

Tschiatschek, S., Stamboliev, E., Schmude, T., Coeckelbergh, M., & Koesten, L. (2024). Challenging the human-in-the-loop in algorithmic decision-making. *arXiv preprint arXiv:2405.10706*.

UNESCO. (2021). *Recommendation on the ethics of artificial intelligence*. Paris: UNESCO.

United Nations. (2015). *Transforming our world: The 2030 agenda for sustainable development*. New York: United Nations.

Van Popering-Verkerk, J., Molenveld, A., Duijn, M., van Leeuwen, C., & Van Buuren, A. (2022). A framework for governance capacity: A broad perspective on steering efforts in society. *Administration & Society, 54*(9), 1767–1794.

van Vugt, M., Colarelli, S. M., & Li, N. P. (2024). Digitally connected, evolutionarily wired: An evolutionary mismatch perspective on digital work. *Organizational Psychology Review, 14*(3), 403–424. https://doi.org/10.1177/20413866241232138

Welzel, C., & Inglehart, R. (2010). Agency, values, and well-being: A human development model. *Social Indicators Research, 97*(1), 43–63. https://doi.org/10.1007/s11205-009-9557-z

Wu, X., Xiao, L., Sun, Y., Zhang, J., Ma, T., & He, L. (2022). A survey of human-in-the-loop for machine learning. *Future Generation Computer Systems, 135*, 364–381.

Xie, Y., Fadahunsi, K. P., Kelleher, C., Tarn, D. M., Grace, A., & O'Donoghue, J. (2025). Towards an inclusive digital health ecosystem. *Bulletin of the World Health Organization, 103*(2), 170–173. https://doi.org/10.2471/BLT.24.292020

Xu, Y., Wang, F., & Zhang, T. (2024). Artificial intelligence is restructuring a new world. *Innovation, 5*(6), 100725. https://doi.org/10.1016/j.xinn.2024.100725

Yatani, K., Sramek, Z., & Yang, C.-L. (2024). AI as extraherics: Fostering higher-order thinking skills in human-AI interaction. *arXiv preprint arXiv:2409.09218*.

Yeung, K., Howes, A., & Pogrebna, G. (2020). AI governance by human rights–centered design, deliberation, and oversight: An end to ethics washing. In *The Oxford handbook of ethics of AI* (pp. 1–25). Oxford University Press. https://doi.org/10.1093/oxfordhb/9780190067397.013.5

Yoder, K. J., & Decety, J. (2018). The neuroscience of morality and social decision-making. *Psychology, Crime & Law, 24*(3), 279–295. https://doi.org/10.1080/1068316X.2017.1414817